\documentclass[letterpaper,12pt]{article}   

\usepackage{amsfonts}
\usepackage{graphicx}
\usepackage{epsfig}
\usepackage{graphics}

\makeatletter
\newcommand{\row}[1]%
{\mathord{\buildrel{\lower3pt%
\hbox{$\scriptscriptstyle\rightarrow$}}\over #1}}
\newcommand{\col}[1]{{#1^{\raisebox{2pt}[\height]%
{$\scriptstyle\downarrow$}}}}
\newcommand{\dyadic}[1]{\mathord{\dyadic@rrow{#1}}}
\newcommand{\dyadic@rrow}[1]{
\begin{picture}(12,12)(-1,0)
\put(-3,12){\makebox(0,0)[t]{$\scriptscriptstyle\downarrow$}}
\put(-3,13){\makebox(0,0)[l]{$\scriptscriptstyle\longrightarrow$}}
\put(5,0){\makebox(0,0)[b]{$#1$}}
\end{picture}
}

\newcommand{\ket}[1]{\bigl| #1 \bigr\rangle}

\input{tcilatex}
\begin{document}
\begin{center}
{\large  Usefulness classes of travelling entangled channels  in
noninertial frames}

\vspace{0.5cm}
N. Metwally\\
Math. Dept., Faculty of Science, South valley university, Aswan,
Egypt.\\
 nasser@Cairo-svu.edu.eg
\end{center}
 \begin{abstract}
The dynamics  of a general   two qubits system in noninetrial
 frame is investigated analytically, where it is assumed that both  of its subsystems  are  differently accelerated.
  Two classes of initial travelling
 states are considered:self transposed  and a generic pure
 states. The entanglement contained in all possible generated entangled
 channels between the qubits and their Anti-qubits is quantified.
The usefulness of the travelling channels  as quantum channels to
perform quantum teleportation is investigated. For the self
transposed classes, we show that  the generalized Werner state is
the most robust class. We show that starting from a class of pure
state, one can generate entangled channels more robust than self
transposed classes.

 \end{abstract}

\section{Introduction}

Due to its importance in quantum information and computation
fields, entanglement has attracted  extensive attentions
\cite{ekert}. The dynamics of entanglement in the real and
practical setting cases  is one of the most promising technique to
develop the efficiency of communication and computations
\cite{Spiller}. Recently, the dynamics of entanglement is extended
to the relativistic world \cite{Als}. In this direction, efforts
have been done to investigate the dynamics of entanglement between
two modes of Dirac modes in noninertial frames. It has been shown
that, the degree of entanglement between entangled observers  is
degraded and consequently the accelerated observer can't obtain
information about his space time \cite{Als}. The dynamics of
entanglement in non-inertial frames for  two qubits system
initially prepared in {\it maximum entangled} state is
investigated in \cite{Als}, where the entanglement is degraded by
the Unruh effect  \cite{un} and its lower limit is enough for
quantum teleportation. The sudden death of entanglement and the
dynamics of mutual information are investigated by Landulfo and
Matsas \cite{andr} in non-inertial frames. The dynamics of the
classical and non-classical correlations of travelling { \it
pseudo-pure }  are investigated by  H. Mehri-Dehnavi et. al.
\cite{Hos}.  The entanglement dynamics of initial  {\it tripartite
class of GHZ} state is investigated in \cite{Jieci1}. Montero et.
al \cite{ Mig}, have introduced {\it tripartite   W -state} as
initial travelling states, where they showed that the entanglement
vanishes completely  in the infinite acceleration limit.

In this contribution, we investigate the dynamics of a general two
qubits state of Dirac field in  non-inertial frame. {\it First},
an analytical solution is introduced  for a general two qubits
system. Particularly, we investigate two classes extensively: the
self transposed class and a class of a generic  pure state. {\it
Second}, the degree of entanglement is quantified for all possible
generated entangled channels between the qubits and their
Anti-qubits. {\it Third}, the usefulness of the travelling
accelerated classes is investigated by means of the fidelity and
the possibility of using them as quantum channels to perform the
quantum teleportation is discussed.

 The outline of the
article is as follows. In Sec.2, we introduce an analytical
solution of the suggested model, where it is  assumed that both
qubits are accelerated. Sec.3 is devoted to investigate the
dynamics of the self transposed classes and a class of a generic
pure state. The  entanglement  of the generated entangled channels
is quantified  in Sec.4. The classification of usefulness
travelling accelerated qubits are  investigated in Sec.5. Finally,
we summarize our results in Sec.6.

\section{The suggested model and its evolution}
Assume that  a source supplies two partners  Alice and Rob with a
general two qubits state which is characterised by $15$
parameters. The general form of this system can be written as,
\begin{equation}\label{qubit1}
\rho_{AR}=\frac{1}{4}(1+\row{s}\cdot\col\sigma_x+\row{t}\cdot\col{\tau}+\row{\sigma}\cdot\dyadic{C}\cdot\col{\tau}),
\end{equation}
where $\row\sigma_i=(\sigma_x,\sigma_y,\sigma_z),
\row\tau_i=(\tau_x,\tau_y,\tau_z)$ are the Pauli matrices for
Alice and Rob's qubit respectively, $\row{s}=(s_x,s_y,s_z)$ with
$s_i=tr\{\rho_{AB} \sigma_i\}$ and $\row{t}=(t_x,t_y,t_z),$
$t_i=tr\{\rho_{AB} \tau_i\}$ are  Bloch vectors for  both  qubits
respectively. The dyadic $\dyadic{C}$is a $3\times 3$ matrix with
elements are defined by $c_{ij}=tr\{\rho_{AB}\sigma_i\tau_i\}$.
For example,
$c_{xx}=tr\{\rho_{AB}\sigma_x\tau_x\},c_{xy}=tr\{\rho_{AB}\sigma_x\tau_y\},
c_{xz}=tr\{\rho_{AB}\sigma_x\tau_x\}$ and so on.  The density
operator (1) $\rho_{AR}$ can be written  as  $4\times 4 $ matrix
its elements are given by,
\begin{eqnarray}\label{cof}
\varrho_{11}&=&\frac{1}{4}(1+s_z+ t_z+c_{zz}), \quad\quad
\varrho_{12}=\frac{1}{4}(t_x- it_y+ c_{zx}+ic_{zy})
\nonumber\\
\varrho_{13}&=&\frac{1}{4}(s_x-is_y+ c_{xz}- ic_{yz}), \quad\
 \varrho_{14}=\frac{1}{4}(c_{xx}- i c_{xy}-ic_{yx}-c_{yy}),
\nonumber\\
\varrho_{21}&=&\frac{1}{4}(t_x- it_y+ c_{zx}+ ic_{zy}), \quad
\varrho_{22}=\frac{1}{4}(1+s_z+ t_z-c_{zz}),
 \nonumber\\
\varrho_{23}&=&\frac{1}{4}(c_{xx}+ i c_{xy}-ic_{yx}+ c_{yy})\quad
\varrho_{24}=\frac{1}{4}(s_x-is_y- c_{xz}+ ic_{yz})
\nonumber\\
 \varrho_{31}&=&\frac{1}{4}(s_x+is_y+ c_{xz}+ ic_{yz}),\quad
 \varrho_{32}= \frac{1}{4}(c_{xx}-i c_{xy}+ic_{yx}+ c_{yy}).
  \nonumber\\
 \varrho_{33}&=&\frac{1}{4}(1-s_z+ t_z-c_{zz}),\quad\quad
 \varrho_{34}=\frac{1}{4}(t_x- it_y-c_{zx}+ ic_{zy}),
 \nonumber\\
 \varrho_{41}&=&\frac{1}{4}(c_{xx}- ic_{xy}+ic_{yx}- c_{yy}),\quad
\varrho_{42}=\frac{1}{4}(s_x+is_y- c_{xz}- ic_{yz}),\quad
\nonumber\\
\varrho_{43}&=&=\frac{1}{4}(t_x+ it_y-c_{zx}- ic_{zy}),\quad
\varrho_{44}=\frac{1}{4}(1-s_z- t_z+c_{zz}).
\end{eqnarray}
The  density operator (\ref{qubit1}) represents a quantum channel
between Alice and Rob in the inertial frame. To study the dynamics
of the travelling channel (\ref{qubit1}) in non-inertial frames we
use what  is called the  Unruh modes \cite{un} which are defined
as,
\begin{eqnarray}\label{Un}
\ket{0}_i&=&\mathcal{C}_i\ket{0}_I\ket{0}_{II}+\mathcal{S}_i\ket{1}_I\ket{1}_{II},
\nonumber\\
\ket{1}_i&=&\ket{1}_{I}\ket{1}_{II},
\end{eqnarray}
where $\mathcal{C}_i=\cos r_i, \mathcal{S}_i=sin r_i,$ with  $tan
r_i=e^{-2\pi\omega_i \frac{c}{a_i}}$, $a_i$ is the acceleration,
$\omega_i$ is the frequency of the travelling qubits, $c$ is the
speed of light and $i=A,R$. In this investigation,  it is assumed
that   Alice and Rob are observers  and can lie in either region
of Rindler space time. Therefore, a uniformly accelerated observer
lying in one wedge of space time is causally disconnected from the
other. This leads to four different situations: (i) the channel
between Alice and Rob $\rho_{\tilde A_I \tilde R_I}$ is in the
region $I$, which requires to trace out the states in mode
$II$.(ii) The channel between Anti-Alice and Anti-Rob
$\rho_{\tilde{A}_{II}\tilde{R}_{II}}$ is in the region $II$ which
requires to trace out the states in mode $I$. (iii) The channel
between Alice and Anti-Rob $\rho_{\tilde{ A}_{I}\tilde {R}_{II}}$,
which is obtained by tracing out Anti-Alice mode in the region
$II$ and the mode of Rob in mode $I$. (v) The channel between Rob
and Anti-Alice, $\rho_{\tilde{A}_{II}\tilde{R}_{I}}$ which is
obtained by tracing  out Alice's mode in the region $I$ and Rob's
mode in the region $II$.

By using Eq.(\ref{qubit1}), (\ref{Un})  and tracing out  the
inaccessible modes in the region $II$, the state of Alice-Rob in
the region $I$ is described by,
\begin{equation}\label{reg1}
\rho_{\tilde{ A}_{I}\tilde{R}_{I}}= \left(
\begin{array}{cccc}
\varrho_{11}\mathcal{C}_1^2\mathcal{C}_2^2&\varrho_{12}\mathcal{C}_1^2\mathcal{C}_2&\varrho_{13}
\mathcal{C}_1\mathcal{C}_2^2&\varrho_{14}\mathcal{C}_1\mathcal{C}_2\\
\\
\varrho_{21}\mathcal{C}_1^2\mathcal{C}_2&\mathcal{C}_1^2(\varrho_{22}+\varrho_{11}\mathcal{S}_2^2)&\varrho_{23}
\mathcal{C}_1\mathcal{C}_2&\varrho_{24}\mathcal{C}_1\\
\\
\varrho_{31}\mathcal{C}_1\mathcal{C}_2&\varrho_{32}\mathcal{C}_1\mathcal{C}_2&
(\varrho_{33}+\varrho_{11}\mathcal{S}_1^2)\mathcal{C}_2^2
&(\varrho_{34}+\varrho_{12}\mathcal{S}_1^2)\mathcal{C}_2\\
\\
\varrho_{41}\mathcal{C}_1\mathcal{C}_2&(\varrho_{42}+\varrho_{31}\mathcal{S}_2^2)\mathcal{C}_1&(\varrho_{43}+
\varrho_{21}\mathcal{S}_1^2)\mathcal{C}_2&\varrho_{44}+\varrho_{33}\mathcal{S}_2^2+(\varrho_{22}+\varrho_{11}\mathcal{S}_2^2)\\
\end{array}
\right).
\end{equation}
Similarly, if we trace out the inaccessible modes in the region
$I$, one gets the density operator of Alice-Rob in the region$II$
as,

\begin{equation}\label{reg2}
\rho_{\tilde{ A}_{II}\tilde{R}_{II}}= \left(
\begin{array}{cccc}
\varrho^{(II)}_{11}&
(\varrho_{43}+\varrho_{21}\mathcal{C}_1^2)\mathcal{S}_2&(\varrho_{42}+\varrho_{31}\mathcal{C}_2^2)
\mathcal{S}_1&\varrho_{41}\mathcal{S}_1\mathcal{S}_2\\
\\
(\varrho_{34}+\mathcal{C}_1^2\varrho_{12})\mathcal{S}_2^2&
(\varrho_{33}+\varrho_{11}\mathcal{C}_1^2)\mathcal{S}_2^2&\varrho_{32}
\mathcal{S}_1\mathcal{S}_2&\varrho_{31}\mathcal{S}_1\mathcal{S}_2^2\\
\\
(\varrho_{24}+\varrho_{13}\mathcal{C}_2^2)\mathcal{S}_2^2&\varrho_{23}\mathcal{S}_1\mathcal{S}_2&
(\varrho_{22}+\varrho_{11}\mathcal{C}_2^2)\mathcal{S}_1^2
&\varrho_{21}\mathcal{S}_1^2\mathcal{S}_2\\
\\
\varrho_{14}\mathcal{S}_1\mathcal{S}_2&\varrho_{13}\mathcal{S}_1\mathcal{S}_2^2&
\varrho_{43}\mathcal{S}_1^2\mathcal{S}_2&\varrho_{11}\mathcal{S}_1^2\mathcal{S}_2^2
\end{array}
\right),
\end{equation}
where
$\varrho^{(II)}_{11}=(\varrho_{22}+\varrho_{11}\mathcal{C}_2^2)\mathcal{C}_1^2+\varrho_{44}+\varrho_{33}\mathcal{C}_2^2$.~

There are two different remaining channels which could be
generated between Alice and Rob.  The first between Alice (in the
first region, I) and Anti-Rob (where Rob  in the second region
II). This state is called Alice-Anti-Rob state, $\rho_{\tilde
{A}_I\tilde{R}_{II}}$ and it is defined as,

\begin{equation}\label{regI-II}
\rho_{\tilde{ A_I}\tilde{R}_{II}}= \left(
\begin{array}{cccc}
(\varrho_{22}+\varrho_{11}\mathcal{C}_2^2)\mathcal{C}_1^2&
\varrho_{21}\mathcal{C}_1^2\mathcal{S}_2^2&(\varrho_{24}+\varrho_{13}\mathcal{C}_2^2)
\mathcal{C}_1&\varrho_{23}\mathcal{C}_1\mathcal{S}_2\\
\\
\varrho_{12}\mathcal{C}_1^2\mathcal{S}_2&
\varrho_{11}\mathcal{C}_1^2\mathcal{S}_2^2&\varrho_{14}
\mathcal{C}_1\mathcal{S}_2&\varrho_{32}\mathcal{C}_1\mathcal{S}_2^2\\
\\
(\varrho_{42}+\varrho_{31}\mathcal{C}_2^2)\mathcal{C}_1&\varrho_{41}\mathcal{C}_1\mathcal{S}_2&
(\varrho_{22}+\varrho_{11}\mathcal{C}_2^2)\mathcal{S}_1^2+(\varrho_{44}+\varrho_{33}\mathcal{C}_2^2)
&(\varrho_{43}+\varrho_{21}\mathcal{S}_1^2\mathcal{S}_2)\\
\\
\varrho_{32}\mathcal{C}_1\mathcal{S}_2&\varrho_{31}\mathcal{C}_1\mathcal{S}_2^2&
(\varrho_{34}+\varrho_{43}\mathcal{S}_1^2)\mathcal{S}_2&(\varrho_{33}+\varrho_{11}\mathcal{S}_1^2)\mathcal{S}_2^2
\end{array}
\right).
\end{equation}
Finally, a similar expression for the density operator between Rob
and Anti-Alice $\rho_{\tilde{R}_I\tilde{A}_{II}}$ is given by,
\begin{equation}\label{regII-I}
\rho_{\tilde{ A_I}\tilde{R}_{II}}= \left(
\begin{array}{cccc}
(\varrho_{33}+\varrho_{11}\mathcal{C}_1^2)\mathcal{C}_2^2&
\varrho_{31}\mathcal{S}_1^2\mathcal{C}_2^2&\varrho_{34}\mathcal{C}_2
\mathcal{S}_1&\varrho_{32}\mathcal{C}_2\mathcal{S}_1\\
\\
\varrho_{13}\mathcal{S}_1^2\mathcal{C}_2^2&
\varrho_{11}\mathcal{S}_1^2\mathcal{C}_2^2&\varrho_{14}
\mathcal{S}_1\mathcal{C}_2&\varrho_{43}\mathcal{C}_2\mathcal{S}_1^2\\
\\
(\varrho_{43}+\varrho_{21}\mathcal{C}_1^2)\mathcal{C}_2&\varrho_{41}\mathcal{S}_1\mathcal{C}_2&
(\varrho_{22}+\varrho_{11}\mathcal{S}_2^2)\mathcal{C}_1^2+(\varrho_{44}+\varrho_{33}\mathcal{S}_2^2)
&(\varrho_{42}+\varrho_{31}\mathcal{S}_2^2)\mathcal{S}_1)\\
\\
\varrho_{23}\mathcal{S}_1\mathcal{C}_2&\varrho_{21}\mathcal{S}_1^2\mathcal{C}_2&
(\varrho_{24}+\varrho_{13}\mathcal{S}_2^2)\mathcal{S}_1&(\varrho_{22}+\varrho_{11}\mathcal{S}_2^2)\mathcal{S}_1^2
\end{array}
\right).
\end{equation}

From Eqs.(\ref{reg1}) and (\ref{reg2}), one can discuss different
cases: the first, if we  set $\mathcal{C}_1=1$ and
$\mathcal{S}_1=0$, then one gets the case where  Rob  stays
stationary while Alice moves with a uniform acceleration. The
second for $\mathcal{C}_2=1$ and $\mathcal{S}_2=0$, one obtains
the case in which Alice stays stationary while Rob moves with  a
uniform acceleration. In the next section  different classes of
initial channels between Alice and Rob  will be considered.

\section{Classes of Entangled Channels}
\begin{enumerate}
\item{ \bf Self transposed class}

A self transposed class is characteristic by zero Bloch vectors
i.e., $\row{s}=\row{t}=0$ and $\rho=\rho^{T}$. The generic form of
the  self transposed class is given by \cite{Englert},
\begin{equation}\label{self}
\rho=\rho^{T}=\frac{1}{4}(1+\row\sigma\cdot\dyadic{C}\cdot\col{\tau}),
\end{equation}
where, the dyadic $\dyadic{C}$ is a $3\times 3$ diagonal matrix
\cite{Englert1}. It has been shown that this class of states is
separable if  $\mathrm{\det}\Bigl\{~\dyadic{C}~\Bigr\}\geq 0 $ or
$\mathrm{tr}\Bigl\{~\dyadic{|C|~}\Bigr\}\leq 1$. However,  if
$\mathrm{\det}\{~\dyadic{C}~\}<0 $ and
$\mathrm{tr}\Bigl\{~\dyadic{|C|~}\Bigr\}>1$, then the sate
(\ref{self}) is entangled and its  degree of entanglement  is
given by the concurrence $\mathcal{C}$ \cite{Woottors},
\begin{equation}
\mathcal{C}=\frac{1}{2}\Bigl(\mathrm{tr}\bigl\{~\dyadic{|C|}~\bigr\}-1\Bigr).
\end{equation}
For this class, one can define three different subclasses:
\begin{enumerate}
\item {\bf  Generalized Werner state}

This class of states sometimes is called $X$ state \cite{Eberly}.
The state (\ref{self}) which represents a two qubits state between
Alice and Rob can be written as,
\begin{equation}\label{Gself}
\rho_{AR}=\frac{1}{4}(1+c_{xx}\sigma_x\tau_x+c_{yy}\sigma_y\tau_y+c_{zz}\sigma_z\tau_z).
\end{equation}
 The non zero elements  of  Eq.(2) are given by,
\begin{eqnarray}\label{cof1}
\varrho_{11}&=&\varrho_{44}=\frac{1}{4}(1+c_{zz}),\quad
\varrho_{22}=\varrho_{33}=\frac{1}{4}(1-c_{zz}),
 \nonumber\\
\varrho_{14}&=&\varrho_{41}=\frac{1}{4}(c_{xx}-c_{yy}),\quad
\varrho_{23}=\varrho_{32}=\frac{1}{4}(c_{xx}+c_{yy}).
\end{eqnarray}
Using Eqs.(\ref{reg1}), (\ref{reg2}) and (\ref{cof1}), on obtains
the density operators in the regions $I$ and $II$ as,
\begin{equation}\label{reg11}
\rho_{\tilde{A}_I\tilde{R}_I}= \left(
\begin{array}{cccc}
\varrho_{11}\mathcal{C}_1^2\mathcal{C}_2^2&0&0&\varrho_{14}\mathcal{C}_1\mathcal{C}_2\\
\\
0&\mathcal{C}_1^2(\varrho_{22}+\varrho_{11}\mathcal{S}_2^2)&\varrho_{23}
\mathcal{C}_1\mathcal{C}_2&0\\
\\
0&\varrho_{32}\mathcal{C}_1\mathcal{C}_2&
(\varrho_{33}+\varrho_{11}\mathcal{S}_1^2)\mathcal{C}_2^2
&0\\
\\
\varrho_{41}\mathcal{C}_1\mathcal{C}_2&0&0&
\varrho_{44}+\varrho_{33}\mathcal{S}_2^2+(\varrho_{22}+\varrho_{11}\mathcal{S}_2^2)\\
\end{array}
\right)
\end{equation}
This state can be written in  the form (1) as
\begin{equation}
\rho_{ \tilde A_I\tilde R_I}=\frac{1}{4}(1+\tilde
{c_{xx}}\sigma_x\tau_x+\tilde {c_{yy}}\sigma_y\tau_y+\tilde
c_{zz}\sigma_z\tau_z)
\end{equation}
where,
\begin{eqnarray}
\tilde{c_{xx}}&=&\mathcal{C}_1\mathcal{C}_2c_{xx},\quad
\tilde{c_{yy}}=\mathcal{C}_1\mathcal{C}_2c_{yy}
\nonumber\\
\tilde{c_{zz}}&=&\frac{1-c_{zz}}{4}(1+
\mathcal{S}_1^2\mathcal{S}_2^2)+\frac{1+c_{zz}}{4}(\mathcal{C}_1^2\mathcal{C}_2^2+\mathcal{S}_1^2\mathcal{S}_1^2-
\mathcal{C}_1^2\mathcal{S}_2^2-\mathcal{S}_1^2).
\end{eqnarray}
Similarly, the density operator in the second region $II$ is given
by,

\begin{equation}\label{reg22}
\rho_{\tilde A_{II}\tilde R_{II}}= \left(
\begin{array}{cccc}
\varrho_{11}&
0&0&\varrho_{41}\mathcal{S}_1\mathcal{S}_2\\
\\
0&(\varrho_{33}+\varrho_{11}\mathcal{C}_1^2)\mathcal{S}_2^2&\varrho_{32}
\mathcal{S}_1\mathcal{S}_2&0\\
\\
0&\varrho_{23}\mathcal{S}_1\mathcal{S}_2&
(\varrho_{22}+\varrho_{11}\mathcal{C}_1^2)\mathcal{S}_1^2
&0\\
\\
\varrho_{14}\mathcal{S}_1\mathcal{S}_2&0&
0&\varrho_{11}\mathcal{S}_1^2\mathcal{S}_2^2
\end{array}
\right),
\end{equation}
By means of  the Bloch vectors and the cross dyadic, the state
(\ref{reg22}) can be written as
\begin{equation}
\rho_{\tilde A_{II}\tilde
R_{II}}=\frac{1}{4}\Bigl[1+\tilde{s}_z^{(II)}\sigma_z+
\tilde{t_z}^{(II)}\tau_z+\tilde{c_{xx}}^{(II)}\sigma_x\tau_x+\tilde{c_{yy}}^{(II)}\sigma_y\tau_y+
\tilde{c_{zz}}^{(II)}\sigma_z\tau_z\Bigr],
\end{equation}
where,
\begin{eqnarray}
\tilde{s_z}^{(II)}&=&\frac{1}{2}(1+\cos2r_1),\quad\quad~\quad
\tilde{t}_z^{(II)}=\frac{1}{2}(1+\cos2r_2),
\nonumber\\
\tilde{C}^{(II)}_{xx}&=&c_{xx}\sin r_1\sin r_2, \quad
\tilde{C}^{(II)}_{yy}=c_{yy}\sin r_1\sin r_2,
\nonumber\\
\tilde{C}^{(II)}_{zz}&=&\frac{1+c_{zz}}{4}\Bigl\{1+\cos
r_1\cos2r_2\Bigl\}
+\frac{1-c_{zz}}{4}\Big\{cos2r_1+\cos2r_2\Bigr\}.
\end{eqnarray}
It is important to point out that, if we assume that  Alice  stays
stationary and Rob moves with  a uniform acceleration $a_R,$ i.e.,
we set $\mathcal{C}_1=1$ and $\mathcal{S}_1=0$ then one gets
$\rho_{A_I\tilde R_{I}}$, which is the same as that obtained by J.
Wang et. al. \cite{Jieci}.

 \item{\bf Bell states}

The class of Bell states  represents an interesting example of the
self transposed states which can be obtained from (\ref{self}) by
setting $c_{xx}=c_{yy}=c_{zz}=\pm 1$. For example, if we set
$c_{xx}=c_{yy}=c_{zz}=-1$ in (\ref{self}), one gets the singlet
state  $\rho_{AR}$ as,
\begin{equation}
\rho_{AR}=\frac{1}{4}(1-\sigma_x\tau_x-\sigma_y\tau_y-\sigma_z\tau_z).
\end{equation}
Now, if we substitute  $c_{ij}=0 ~\mbox{for}~ i \neq j$ in
(\ref{cof}), then the non zero elements are,
\begin{equation}\label{cof2}
\varrho_{11}=\varrho_{33}=\frac{1}{2} ,
\varrho_{23}=\varrho_{32}=-\frac{1}{2}.
\end{equation}
By using (\ref{reg1}), (\ref{reg2}) and (\ref{cof2}) one gets the
density operators for the two qubits in the region $I$,
$\rho_{\tilde{ A}_I\tilde{R}_I}$ and in the region $II$,
$\rho_{\tilde{ A}_{II}\tilde{R}_{II}}$  as,

\begin{eqnarray}\label{reg-1}
\rho_{\tilde{ A}_{I}\tilde{R}_{I}}&=&\frac{1}{4}\Bigl\{1-\cos
r_1\cos r_2(\sigma_x\tau_x+\sigma_y\tau_y)
-\frac{1}{2}(1+\cos2r_1\cos2 r_2)\sigma_z\tau_z\Bigr\},
\nonumber\\
\rho_{\tilde{ A}_{II}\tilde{R}_{II}}&=&\frac{1}{4}\Bigl\{1-\sin
r_1\sin r_2(\sigma_x\tau_x+\sigma_y\tau_y)
+\frac{1}{2}(\cos2r_1+\cos2 r_2)\sigma_z\tau_z\Bigr\},
\end{eqnarray}
where it is assumed that  both  qubits are accelerated. However,
if we assume that Alice  stays stationary and Rob is accelerated
then the states(\ref{reg-1}) reduce to,
\begin{eqnarray}
\rho_{A_I\tilde{R}_I}&=&\frac{1}{4}\Bigl\{1-\cos
r_2(\sigma_x\tau_x+\sigma_y\tau_y) -\frac{1}{2}(1+cos2
r_2)\sigma_z\tau_z\Bigr\},
\nonumber\\
\rho_{ A_{II}\tilde{R}_{II}}&=&\frac{1}{4}\Bigl\{1
+\frac{1}{2}(1+cos2 r_2)\sigma_z\tau_z\Bigr\},
\end{eqnarray}
where we set $\mathcal{C}_1=0$ and $\mathcal {S}_1=0$ in
(\ref{reg-1}).

\item{\bf   Werner State }

This class of states is defined as \cite{Wer},
\begin{equation}
\rho_{w}=\frac{1}{4}(1+x\row\sigma\cdot\dyadic{O}_{en}\cdot\col{\tau}),
\end{equation}
where $\dyadic{O}_{en}$ is unimodular, orthogonal en-dyadic
\cite{Englert1}.  It has been shown that this state is separable
for $x\in[-\frac{1} {3} ,\frac{1}{3} ]$ and nonseparable for
$\frac{1} {3} < x <1$.
 In this case  the non zero elements of Eq.(2)  are given
 by,
\begin{equation}\label{cofWR}
\varrho_{11}=\varrho_{33}=\frac{1+x}{4}, \quad
\varrho_{22}=\varrho_{44}=\frac{1-x}{4},\quad
\varrho_{23}=\varrho_{32}=\frac{x}{2}.
\end{equation}
 By using  (\ref{reg1}) and (\ref{cofWR}),  the
density operators in the regions $I$  and $II$ take the form,
\begin{equation}
\rho_{\tilde{ A}_i\tilde{ R}_i}=\frac{1}{4}\Bigl\{1+\tilde
s_z^{(i)}\sigma_x+\tilde t_z^{(i)}\tau_z+\tilde
{c}_{xx}^{(i)}\sigma_x\tau_x+\tilde
{c}_{yy}^{(i)}\sigma_x\tau_x+\tilde
{c}_{zz}^{(i)}\sigma_x\tau_x\Bigr\},\quad i=I, II
\end{equation}
where,
\begin{eqnarray}
\tilde
s_z^{(I)}&=&\frac{1}{4}\Bigl\{(1+x)\cos2r_1-(1-x)(1-\cos2r_1)\Bigr\},
\nonumber\\
\tilde
t_z^{(I)}&=&\frac{1}{4}\Bigl\{(1+x)\cos2r_2-(1-x)(1-\cos2r_2)\Bigr\},
\nonumber\\
\tilde{c}_{xx}^{(I)}&=&\tilde{c}_{yy}=x\cos r_1\cos r_2,
\nonumber\\
\tilde{c}_{zz}^{(I)}&=&\frac{1}{2}\Bigl\{(1+x)\cos 2r_1\cos
2r_2-(1-x)(\cos 2r_1+\cos 2r_2)\Big\},
\end{eqnarray}
for the density operator in the region $I$ and
\begin{eqnarray} \tilde
s_z^{(II)}&=&\frac{1}{2}\Bigl\{1+\cos 2r_1\Bigr\},
\nonumber\\
\tilde t_z^{(II)}&=&\frac{1}{2}\Bigl\{1+\cos 2r_2\Bigr\},
\nonumber\\
\tilde{c}_{xx}^{(II)}&=&\tilde{c}_{yy}^{(II)}=x\sin r_1\sin r_2,
\nonumber\\
\tilde{c}_{zz}^{(II)}&=&\frac{1}{4}\Bigl\{(1+x)(1+\cos 2r_1\cos
2r_2)+(1-x)(\cos 2r_1+\cos 2r_2)\Big\},
\end{eqnarray}
for the density operator in the second region $II$.
\end{enumerate}

 \item{\bf Generic Pure state}

This class  is characteristed by  one parameter $ p$, which is
equal to the  length of the Bloch vectors i.e $|\row s|=|\row
t|=p$ \cite{Englert1}. This type of pure states  can be written
as,
\begin{equation}\label{Pure}
\rho_{\mathrm{p}}=\frac{1}{4}\Bigl(1+p(\sigma_x-\tau_x)-\sigma_x\tau_x-\sqrt{1-p^2}(\sigma_y\tau_y+\sigma_z\tau_z)\Bigr),
\end{equation}
where  Bloch vectors and the non-zero elements of the  cross
dyadic are given by,
\begin{equation}
\row{s}=(p,0,0),\quad~ \row{t}=(-p,0,0),\quad~c_{xx}=-1,\quad~
c_{yy}=c_{zz}=-\sqrt{1-p^2},
\end{equation}
and the non zero elements (2) of the density operator  (27) are
given by,
\begin{eqnarray}\label{CoPur}
\varrho_{11}&=&\frac{1-q}{4}=\varrho_{44}=-\varrho_{23}=-\varrho_{32},\quad
\varrho_{22}=\frac{1+q}{4}=\varrho_{44}=-\varrho_{14}=-\varrho_{41},
\nonumber\\
\varrho_{13}&=&\varrho_{24}=\varrho_{31}=\varrho_{42}=\frac{p}{4}=
-\varrho_{12}=-\varrho_{21}=-\varrho_{34}=-\varrho_{43},
\end{eqnarray}
where $q=\sqrt{1-p^2}$. Using  (\ref{reg1}),(\ref{reg2}) and
(\ref{CoPur}), one obtains the density operator in the regions $I$
and $II$ respectively as,
\begin{equation}\label{Pure}
\rho_{\tilde {A}_i\tilde
{R}_i}=\frac{1}{4}\Bigl(1+s_x^{(i)}\sigma_x+s_z^{(i)}\sigma_z+t_x^{(i)}\tau_x+t_z^{(i)}\tau_z-
\tilde
c_{xx}^{(i)}\sigma_x\tau_x+c_{xz}^{(i)}\sigma_x\tau_z+c_{zx}^{(i)}\sigma_z\tau_x+\tilde
c_{yy}^{(i)}\sigma_y\tau_y+\tilde
c_{zz}^{(i)}\sigma_z\tau_z\Bigr),
\end{equation}
where $i=I,II$ for the  first and second regions respectively. In
the region $I$, the density operator between Alice and Rob
$\rho_{\tilde A_I\tilde R_{I}}$ is characteristic by,
\begin{eqnarray}
\tilde s_x^{(I)}&=&p~ \mathcal{C}_1,\quad \tilde
s_z^{(I)}=\frac{1}{2}(\cos2r_1-1), \quad \tilde t_x^{(I)}=-p~
\mathcal{C}_2,\quad \tilde t_z^{(I)}=\frac{1}{2}(\cos2r_2-1),
\nonumber\\
\tilde c_{xx}^{(i)}&=&-\mathcal{C}_1\mathcal{C}_2, \quad \tilde
c_{xz}^{(I)}=-\frac{p}{2} \mathcal{C}_1(1-\cos2r_2),\quad \tilde
c_{zx}^{(I)}=-\frac{p}{2} \mathcal{C}_2(1-\cos2r_1), \nonumber\\
c_{yy}^{(I)}&=&-q~\mathcal{C}_1\mathcal{C}_2,\quad
c_{zz}^{(I)}=-\frac{q}{2}\cos2r_1+\frac{1}{4}\Bigl\{(1-q)-(1+q)\cos2r_2\Bigr\},.
\end{eqnarray}
and for the second region II, the density operator $\rho_{\tilde
A_{II}\tilde R_{II}}$ is described by,

\begin{eqnarray}
\tilde s_x^{(II)}&=&p ~\mathcal{S}_1,\quad \tilde
s_z^{(II)}=\frac{1}{2}(1+\cos2r_1), \quad \tilde t_x^{(II)}=-p
\mathcal{S}_2,\quad \tilde t_z^{(II)}=\frac{1}{2}(1+\cos2r_2),
\nonumber\\
\tilde c_{xx}^{(II)}&=&-\mathcal{S}_1\mathcal{S}_2, \quad \tilde
c_{xz}^{(II)}=-\frac{p}{2} \mathcal{S}_1(1+\cos2r_2),\quad \tilde
c_{zx}^{(II)}=-\frac{p}{2} \mathcal{S}_2(1-\cos2r_1), \nonumber\\
c_{yy}^{(II)}&=&-q\mathcal{S}_1\mathcal{S}_2,\quad
c_{zz}^{(II)}=-\frac{1-q}{4}(1+\cos2r_1\cos2r_2)+\frac{1+q}{4}(\cos2r_1+\cos2r_2).
\nonumber\\
\end{eqnarray}

\end{enumerate}

\section{Entanglement}
\begin{figure}[b!]\label{Fig12Acc}
  \begin{center}
  \includegraphics[width=19pc,height=14pc]{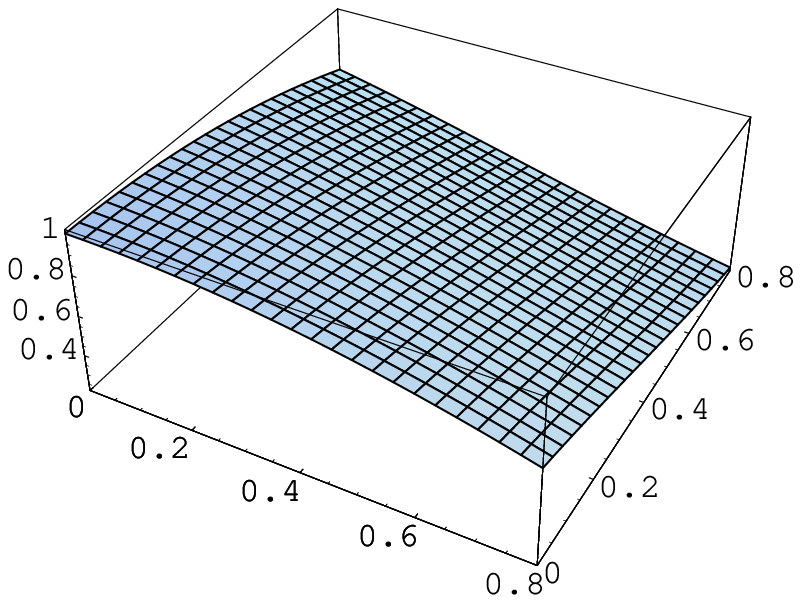}~
   \includegraphics[width=19pc,height=14pc]{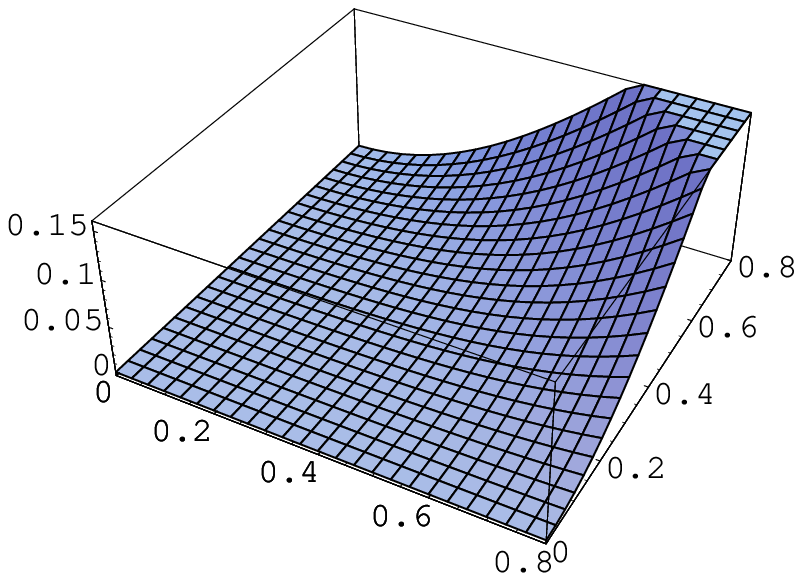}
    \put(-470,80){$\mathcal{C}$}
 \put(-230,80){$\mathcal{C}$}
   \put(-263,150){$(a)$}
 \put(-28,150){$(b)$}
 \put(-150,20){$r_a$} \put(-250,50){$r_b$}
 \put(-400,20){$r_a$}
\put(-20,50){$r_b$}\\
  \includegraphics[width=19pc,height=14pc]{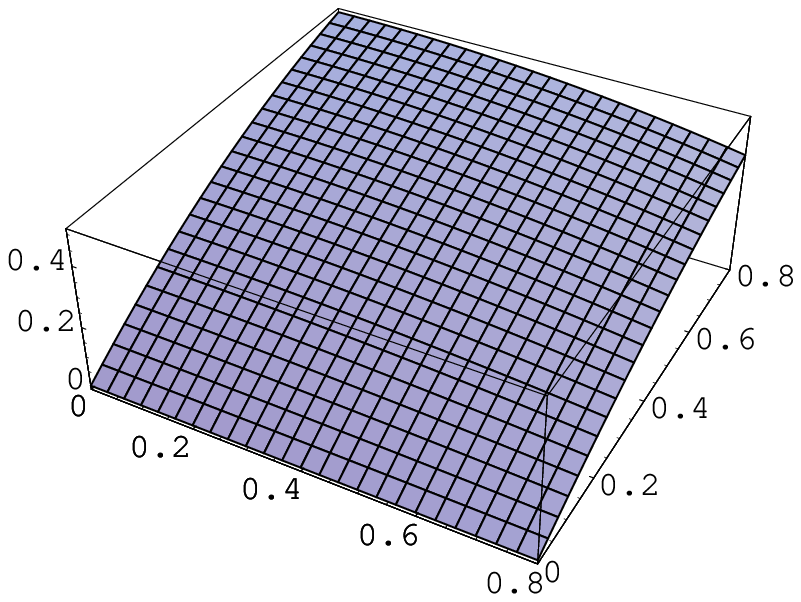}
   \includegraphics[width=19pc,height=14pc]{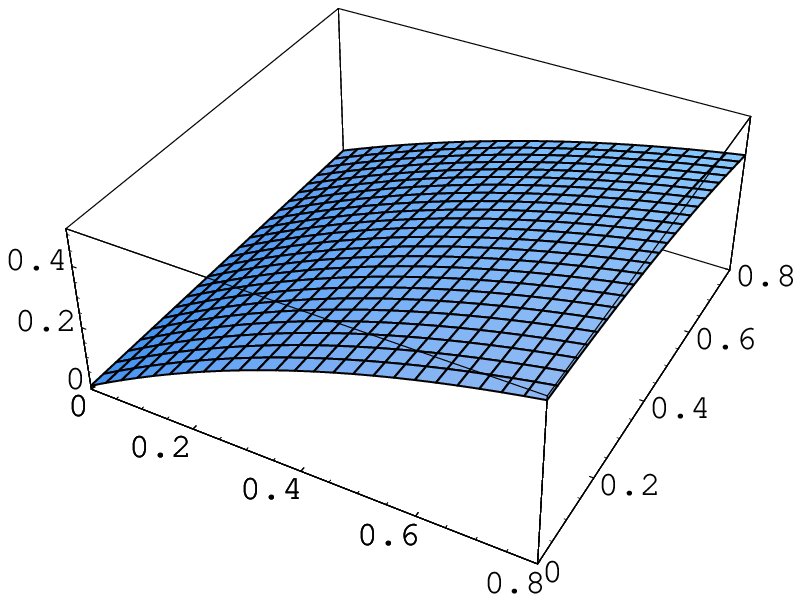}
 \put(-470,80){$\mathcal{C}$}
 \put(-230,80){$\mathcal{C}$}
\put(-150,20){$r_a$} \put(-250,50){$r_b$}
 \put(-400,20){$r_a$}
\put(-20,50){$r_b$}
 \put(-263,150){$(c)$}
 \put(-28,150){$(d)$}
     \caption{The  entanglement of the generated entangled channels from a system  initially prepared
     in maximum entangled
    state (MES)  between:(a) Alice and Rob in the first region
    $\rho_{\tilde{A}_I\tilde{R}_I}$,
    (b) Alice and Rob in the second region
    $\rho_{\tilde{A}_{II}\tilde{R}_{II}}$,
    (c)  Alice and Anti-Rob $\rho_{\tilde{A}_I\tilde{R}_{II}}$, (d) Rob and anti-Alice, $\rho_{\tilde{R}_I\tilde{A}_{II}}$.}
  \end{center}
\end{figure}

In this section, we investigate the entanglement  behavior for
different classes  of initial states settings. The earliest
investigation has considered only one qubit moving with a uniform
acceleration while the other one stays stationary \cite{Jieci1}.
In the current study, we investigate extensively all different
situations.

To measure the  entanglement  of the generated entangled channels
between the qubits and their Anti-qubits, we use
Wotters'concurrence \cite{Hill},

\begin{equation}
\mathcal{C}=\max\Bigl\{0,\lambda_1-\lambda_2-\lambda_3-\lambda_4\Bigr\},
\end{equation}
where $\lambda_1\geq\lambda_2\geq\lambda_3\geq\lambda_4$ and
$\lambda_i$ are the eigenvalues of the density operator
$\rho=\sigma_y\tau_y\rho^*\sigma_y\tau_y$, $\rho^*$ is the complex
conjugate of $\rho$.

Fig.(1), shows the dynamics of the concurrence $\mathcal{C}$  for
a system  initially prepared in maximum entangled state.  In this
investigation we assume that both qubits are equally accelerated
i.e., $r_a=r_b=r$. Fig.(1a) displays the dynamics of the
concurrence $\mathcal{C}$ of the generated entangled channel
between Alice and Rob in the first region $\rho_{\tilde A_I\tilde
R_I}$. It is clear that, since we start with MES ,the concurrence
$\mathcal{C}=1$ at $r_a=r_b=0$. However if, the first qubit is
accelerated while the  second qubit  stays stationary i.e.,
$r_a=r$ and $r_b=0$, then the concurrence decreases smoothly and
gradually and does'nt vanish even when the acceleration tends to
infinity. However, if the second qubit is  also accelerated, then
the concurrence decreases faster and vanishes completely at
infinity.

Fig.(1b) displays the concurrence dynamics  of the generated
channel between Alice and Rob in the second region $\rho_{\tilde
{A}_{II}\tilde{R}_{II}}$. It is clear that, the system is
disentangled for small values of the accelerations. However an
entangled channel is generated between Alice and Rob in the second
region with  $\mathcal{C}\leq0.15$.

In Fig.(1c), we quantify the degree of entanglement which is
generated between Alice (in the first region $I$) and Anti-Rob (in
the second region $II$).  This figure displays that the maximum
value of $\mathcal{C}=0.4$, is reached for zero acceleration of
both qubits. However if only one qubit is accelerated, then the
evolved channel  between Alice and Rob becomes separable. As soon
as the second qubit is accelerated, an entangled channel is
generated between Alice and Anti-Rob, $\rho_{\tilde
A_{I}\tilde{R}_{II}}$. The concurrence of this channel  increases
as the acceleration of Rob's particle increases. The concurrence
of the  generated channel between Rob and Anti-Alice is depicted
in Fig.(1d).  It is clear that, this channel is separable for zero
accelerations. However as soon as  the acceleration of Alice
increases  an entangled channel  is generated. The maximum value
of the concurrence of this channel is $\mathcal{C}\leq 0.4$.

Fig.(2)  displays the dynamics of the concurrence $\mathcal{C}$
for a Werner class and a generalized Werner state. Fig.(2a)
describes the concurrence dynamics of the generated channel
between Alice and Rob in the first region,
$\rho_{\tilde{A}_I\tilde{R}_I}$ where it is assumed that both
qubits are accelerated.  At zero accelerations ($r_a=r_b=0$), the
concurrence depends on the initial degree of entanglement. The
entanglement decays smoothly and gradually as soon as any qubit is
accelerated.  However, the entanglement decays faster when both
qubits  are accelerated and completely vanishes at infinite
accelerations. Fig.(2b) shows the behavior of the entanglement for
the generalized Werner state
 where we set $c_x=0.7,c_y=0.5$ and $c_z=0.3$. This figure
displays a similar behavior as that shown in Fig.(2a) at zero
acceleration. For large accelerations the entanglement decreases
suddenly to reach its minimum values. Theses  minimum values
increase for larger accelerations of both qubits.

\begin{figure}
  \begin{center}
   \includegraphics[width=19pc,height=14pc]{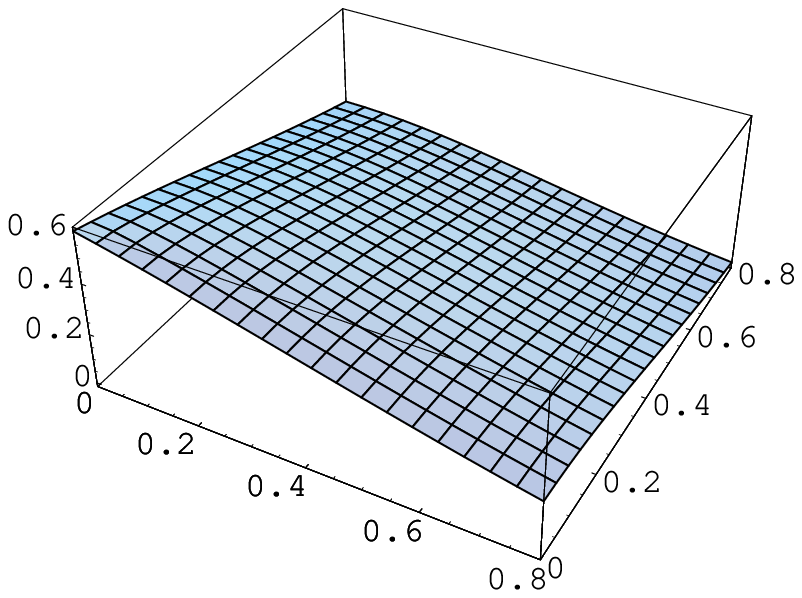}
 \includegraphics[width=19pc,height=14pc]{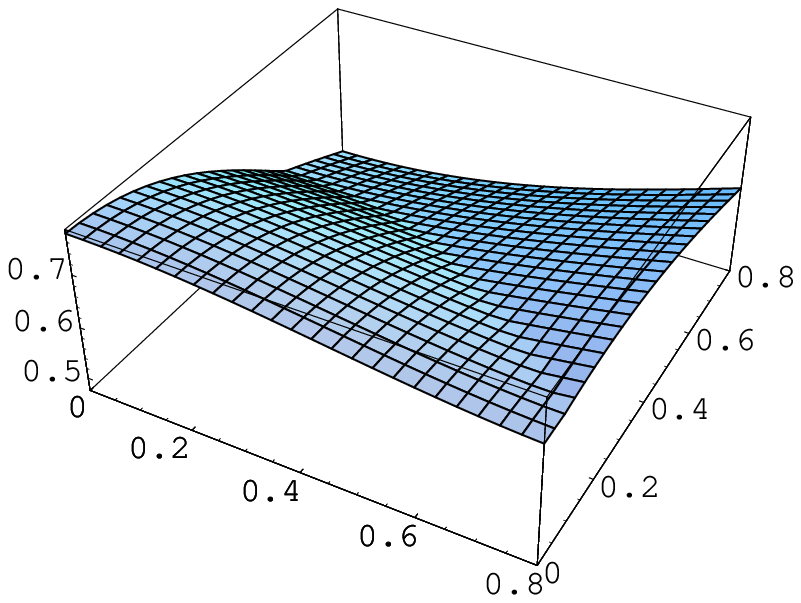}
 \put(-470,80){$\mathcal{C}$}
 \put(-230,80){$\mathcal{C}$}
\put(-150,20){$r_a$} \put(-250,50){$r_b$}
 \put(-400,20){$r_b$}
\put(-20,50){$r_b$}
 \put(-263,150){$(a)$}
 \put(-28,150){$(b)$}
     \caption{The entanglement of the generated  state between Alice and Rob, $\rho_{A_IR_I}$for (a) Werner state with
     $x=0.6$, (b) generalized Werner state with $c_x=0.7,c_y=0.5,c_z=0.3$. }
  \end{center}
\end{figure}

Fig.(3), describes  the concurrence  dynamics of a travelling
state   initially prepared  in  the generic pure state (29),where
 we consider  that both qubits are accelerated with the same
acceleration.   It is clear that at  $p=0$, namely, the initial
system is MES, the concurrence is maximum $(\mathcal{C}=1)$. As
the accelerations of both qubits increase, the concurrence
decreases gradually. For larger values of $p$ (i.e. the initial
system is partially entangled), the concurrence decreases
gradually and completely vanishes at $p=1$.

From Figs.(1) and (2), one can conclude some important results:
{\it first} the  entanglement of the generated channels depends on
the initial entanglement of the travelling state. {\it Second},
the generated channel in the second region behaves classically and
non-classically for small range of finite accelerations. {\it
Third}, the  entanglement between one qubit and the Anti-of the
second qubit depends on the acceleration of the Anti particle.
{\it  Fourth}, the generalized Werner state is the most robust
type of the self transposed classes. {\it Fifth}, among all theses
classes  the generic pure state is more robust than the self
transposed classes even the travelling sate is maximum entangled.

\begin{figure}\label{Fig12Acc}
  \begin{center}
  \includegraphics[width=19pc,height=14pc]{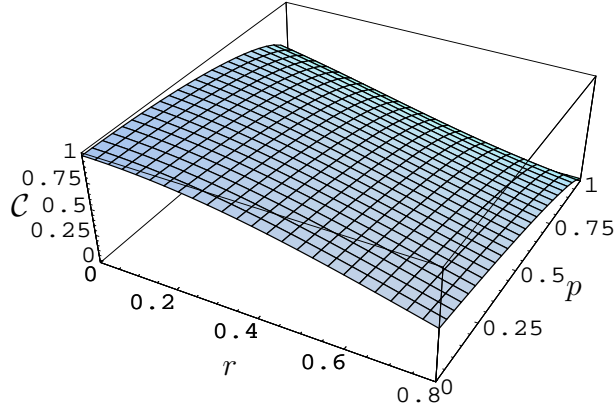}
\put(-230,80){$\mathcal{C}$} \put(-150,20){$r$} \put(-20,50){$p$}
    \caption{(a) The degree of entanglement of a system is initially prepared in a generic pure
    state,  where it is assumed that
     the two qubits are accelerated i.e. $r_a=r_b=r$}
  \end{center}
\end{figure}

\section{Usefulness classes}

In this section, we  investigate the usefulness of the different
classes from two different points of view. First, we quantify the
fidelity
$\mathcal{F}=tr\Bigl\{\rho_{\mathrm{final}}\rho_{\mathrm{initial}}\Bigr\}$
 of the travelling channel in the different regions \cite{Ping}.
Second we investigate the possibility of using the travelling
channels to perform the original
 quantum teleportation \cite{Bennt}.
\begin{enumerate}
\item{ \bf Fidelities of the travelling states}

 Let us first consider that the travelling channel
is  prepared initially in a class of self transposed states
(\ref{self}). For this class, the fidelity $\mathcal{F}_I$ of the
travelling state in the first region is given by,
\begin{eqnarray}
\mathcal{F}_I&=&\frac{1+C_{zz}}{4}\Bigl\{\varrho_{11}\mathcal{C}_1^2\mathcal{C}_2^2+\mathcal{\varrho}_{44}\Bigr\}+
\frac{C_{xx}+c_{yy}}{2}(\varrho_{23}+\varrho_{32})\mathcal{C}_1\mathcal{C}_2
\nonumber\\
&+&\frac{C_{xx}-C_{yy}}{4}(\varrho_{14}+\varrho_{41})\mathcal{C}_1\mathcal{C}_2+
\frac{1-C_{zz}}{4}\Bigl\{\mathcal{C}_1^2(\varrho_{22}+\varrho_{11}\mathcal{S}_2^2)+
\mathcal{C}_2^2(\varrho_{33}+\varrho_{11}\mathcal{S}_1^2)\Bigr\},
\nonumber\\
\end{eqnarray}
while, the fidelity of the travelling channel in the region $II$
is,
\begin{eqnarray}
\mathcal{F}_{II}&=&\frac{1+C_{zz}}{4}\Bigl\{\varrho_{11}\mathcal{S}_1^2\mathcal{S}_2^2+\mathcal{\varrho}_{11}\Bigr\}+
\frac{C_{xx}+c_{yy}}{2}(\varrho_{23}+\varrho_{32})\mathcal{S}_1\mathcal{S}_2
\nonumber\\
&&+\frac{C_{xx}-C_{yy}}{4}(\varrho_{14}+\varrho_{41})\mathcal{S}_1\mathcal{S}_2
+\frac{1-C_{zz}}{4}\Bigl\{\mathcal{S}_2^2(\varrho_{33}+\varrho_{11}\mathcal{C}_1^2)+
\mathcal{S}_1^2(\varrho_{22}+\varrho_{11}\mathcal{C}_1^2)\Bigr\}.
\nonumber\\
\end{eqnarray}
 Similarly, one can obtain a form of the fidelity of the channels
 between Alice and Anti-Rob, $\rho_{\tilde{A}_I\tilde{R}_{II}}$ and between
 Rob, Anti-Alice $\rho_{\tilde{R}_{I}\tilde{A}_{II}}$.
\begin{figure}[t!]
  \begin{center}\label{fid}
    \includegraphics[width=19pc,height=14pc]{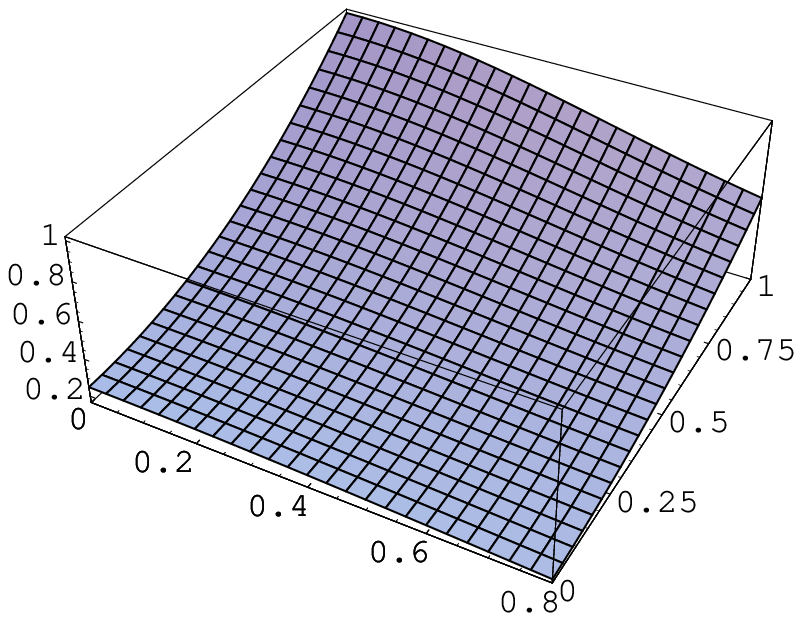}~\quad
    \includegraphics[width=19pc,height=14pc]{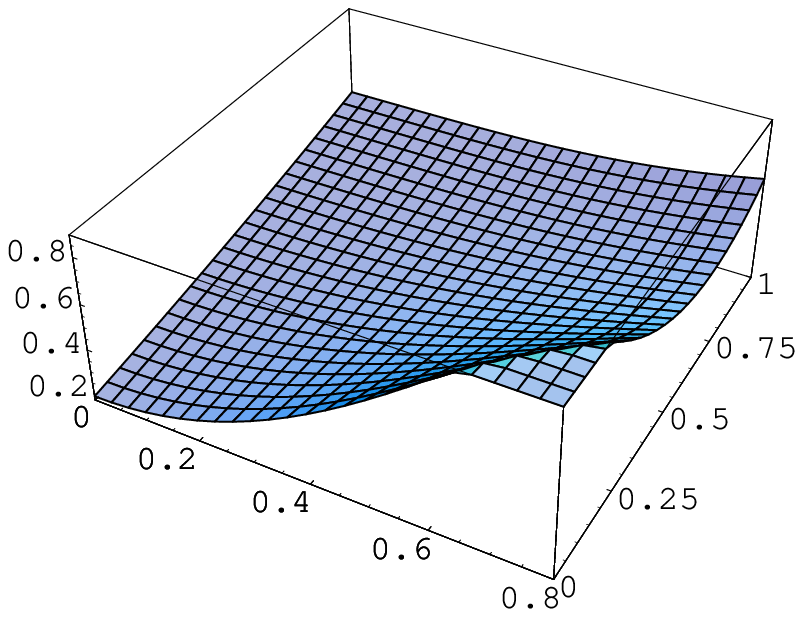}
 \put(-480,80){$\mathcal{F}_I$}
 \put(-240,80){$\mathcal{F}_{II}$}
\put(-150,20){$r$} \put(-260,50){$x$}
 \put(-400,20){$r$}
\put(-20,50){$x$}
 \put(-420,150){$(a)$}
 \put(-28,150){$(b)$}\\
 \includegraphics[width=19pc,height=14pc]{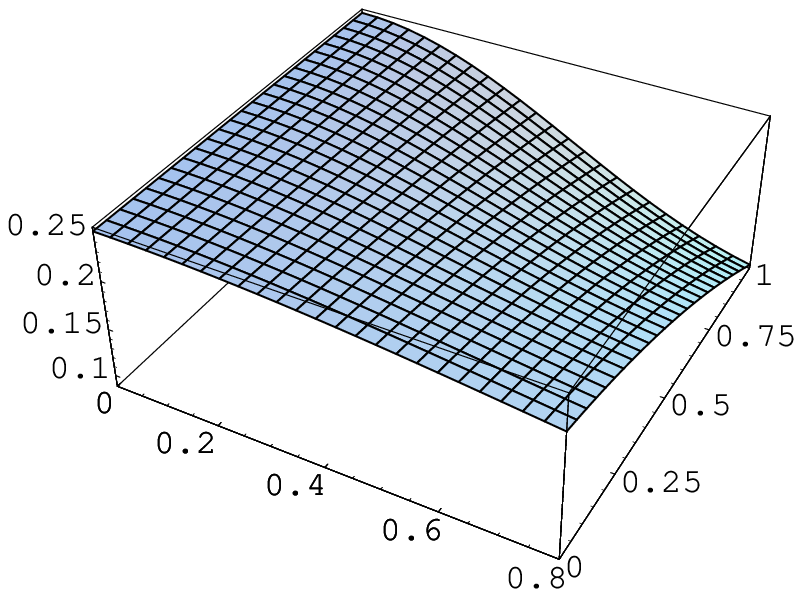}~\quad
 \includegraphics[width=19pc,height=14pc]{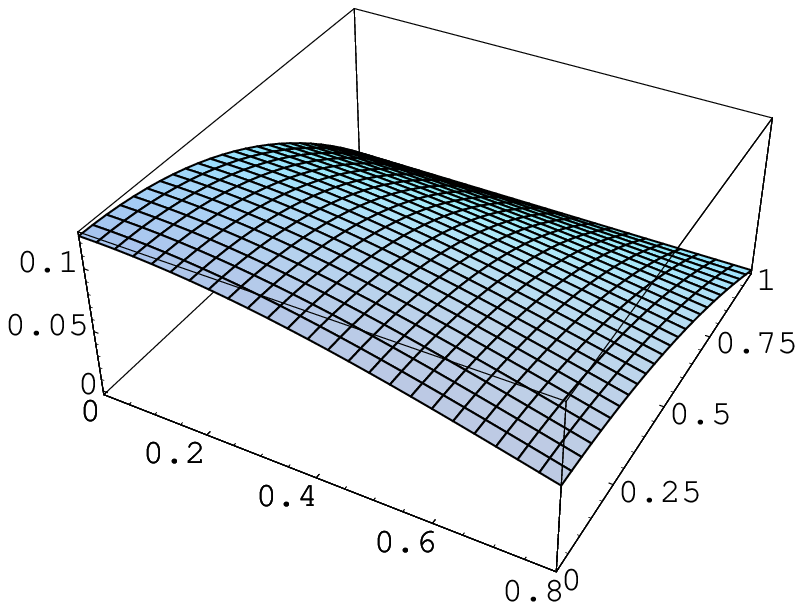}
 \put(-263,150){$(c)$}
 \put(-28,150){$(d)$}
  \put(-400,20){$r$}
\put(-20,50){$x$} \put(-150,20){$r$} \put(-260,50){$x$}
\put(-485,88){$\mathcal{F}_{\tilde{A}_I\tilde{R}_{II}}$}
 \put(-245,88){$\mathcal{F}_{\tilde{R}_I\tilde{A}_{II}}$}
    \caption{The Fidelity of the  travelling state  between: (a) Alice and Rob,
    $\rho_{\tilde A_I\tilde R_I}$,
    (b) Alice and Rob in the second region $\rho_{A_{II}R_{II}}$,
    (c) Alice and Anti-Rob, $\rho_{\tilde{A}_I\tilde{R}_{II}}$,
    (d) Rob and Anti-Alice, $\rho_{\tilde{R}_I\tilde{A}_{II}}$,
    where it is assumed that both particles are accelerated i..e $r_a=r_b=r$ for a system  initially prepared in
   a self transposed type(MES or Werner).}
  \end{center}
\end{figure}
The dynamics of the fidelities is described in Fig.(4), where it
is assumed that both qubits are accelerated with the same
acceleration i.e. $r_a=r_b=r$. This class of state represents the
maximum entangled state($x=1$) and different classes of Werner
states for ($0<x<1$). Fig.(4a) displays the dynamics of
$\mathcal{F}_I$ for different classes of self transposed states.
Let us consider the {\it first} class which is described by
$c_{xx}=c_{yy}=c_{zz}=x$ where  $x\in[0,0.20]$. For this class,
the fidelity $\mathcal{F}\in[0,0.2]$, where the largest value is
reached  at $r=0$ and the smallest value is reached at infinite
acceleration i.e $r=\frac{\pi}{4}$. For the {\it second }class
which includes all classes with  $x\in(0.2,1.0)$, the fidelity
$\mathcal{F}_I$, increases for small values of $r$ and larger
values of $x$. For the second class the fidelity is better than
the first one even for larger  values of $r$. Finally for  the
{\it third} class with $x=1$, which represents the maximum
entangled class, the fidelity $\mathcal{F}$ is maximum for $r=0$
and $x=1$. However as $r$ increases the fidelity of the travelling
state is the best one, where $\mathcal{F}\in[0.8,1]$.

Fig.(4b) shows the behavior of the fidelity in the second region
$\mathcal{F}_{II}$ for different classes of the self transposed
states which are characteristic by the parameter $x$. The fidelity
is almost zero for small values of $r$ and  $x$, i.e. initially
the travelling states have a smaller degree of entanglement. As
$x$ increases, the maximum  fidelity $\mathcal{F}_{II} \in[0,0.4]$
and increases slightly for larger values  $(x=1)$.

The fidelity of the generated channel  between Alice and Anti-Rob,
$\rho_{\tilde{A}_I,\tilde{R}_{II}}$ is displayed in Fig.(4c). It
is clear that the fidelity decreases as one increases the
acceleration of both qubits. However for larger values of the
parameter $x$, the fidelity decreases faster and almost vanishes
for maximum entangled state. i.e., at $x=1$. The dynamics of the
fidelity between Rob and ant-Alice is depicted in Fig.(4d).  The
behavior of the fidelity is similar to that shown in Fig.(4c), but
the maximum fidelity is smaller and decreases faster for larger
values of the accelerations and the parameter $x$.

Now, we consider the dynamics of the fidelity for a system
initially prepared in a generic pure state (29). For this class,
the fidelity of the travelling channel in the first region $I$ is
given by,
\begin{equation}
\mathcal{F}_{I}^{\mathrm{p}}=(\frac{1-q}{4})^2\left\{\mathcal{C}_1^2\mathcal{C}_2+2\mathcal{C}_1\mathcal{C}_2+
\mathcal{S}_1^2\mathcal{S}_2^2+1\right\}+(\frac{1+q}{4})^2(\mathcal{C}_1+\mathcal{C}_2)^2+
(\frac{p}{4})^2(4\mathcal{C}_2+2\mathcal{C}_1),
\end{equation}
while in the second region $II$, it is given by,
\begin{eqnarray}
\mathcal{F}^{\mathrm{p}}_{II}&=&\left(\frac{1+q}{4}\right)^2\Bigl[1+2\mathcal{S}_1\mathcal{S}_2+\mathcal{C}_1^2\mathcal{C}_2^2
+\mathcal{S}_1^2\mathcal{S}_2^2\Bigr]+\left(\frac{1+q}{4}\right)^2(\mathcal{S}_1+\mathcal{S}_2)^2++\frac{p^2}{4}(\mathcal{S}_1+\mathcal{S}_2)
\nonumber\\
&+&\frac{1-q^2}{16}\left(\mathcal{C}_1^2+\mathcal{C}_2^2+\mathcal{C}_1^2\mathcal{S}_2^2+\mathcal{S}_1^2\mathcal{C}_2^2\right).
\end{eqnarray}
\begin{figure}\label{Figfidp}
  \begin{center}
    \includegraphics[width=19pc,height=14pc]{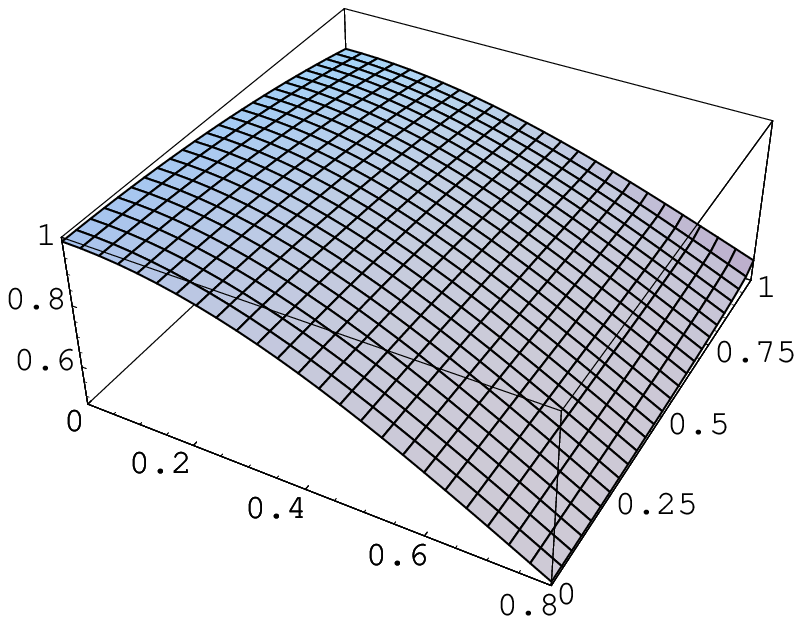}~\quad
 \includegraphics[width=19pc,height=14pc]{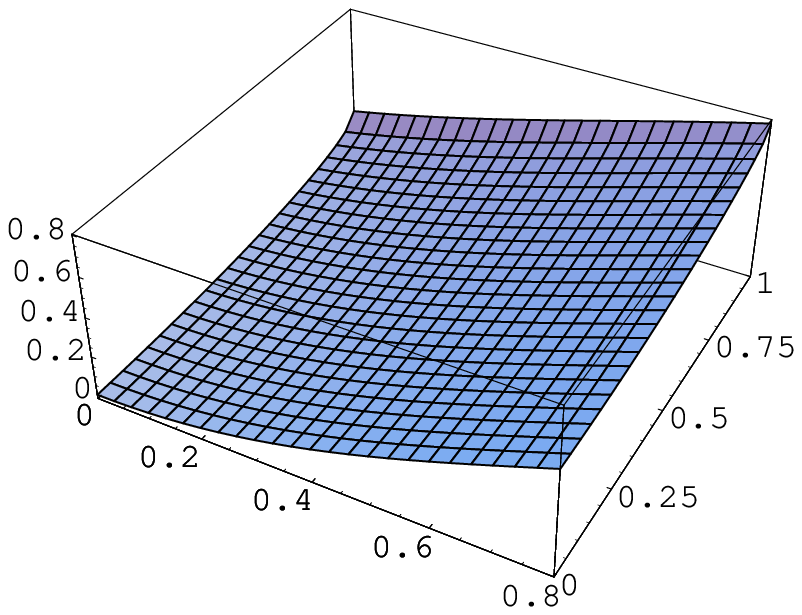}
 \put(-485,80){$\mathcal{F}^{\mathrm{p}}_I$}
 \put(-240,80){$\mathcal{F}^{\mathrm{p}}_{II}$}
\put(-150,20){$r$} \put(-260,50){$p$}
 \put(-400,20){$r$}
\put(-20,50){$p$}
 \put(-420,150){$(a)$}
 \put(-28,150){$(b)$}
    \caption{The Fidelity of the  travelling state ,
    where it is assumed that both particles are accelerated i.e, $r_a=r_b=r$ for a system  initially prepared in
   a pure state,(a) for the first region $I$ and (b) for the second region $II$.}
  \end{center}
\end{figure}

The dynamic of the fidelity for a system  initially prepared
 in the generic pure state (29) is displayed in Fig.(5).
In the first region, the fidelity  $\mathcal{F}_I$ is shown in
Fig.(5a). It is clear that at $p=0$ i.e.,  the travelling  state
is maximum entangled channel ($\mathcal{F}^{\mathrm{p}}_I=1$)  at
$r=0$. For larger $p$ and smaller $r$, the fidelity  decreases
slightly. However the fidelity decreases smoothly and gradually
for larger values of $r$ and completely vanishes at infinity.
Fig.(5b) describes the behavior of the fidelity of the travelling
state in the second region $\mathcal{F}^{\mathrm{p}}_{II}$. For
small values of $r$ and even larger values of $p$, the fidelity is
very small. However as $r$ increases, the fidelity increases
smoothly and reaches its upper bound at $p=1$.

\item{\bf Usefulness  classes for teleportation}
\begin{figure}[t!]\label{Fig:tp}
  \begin{center}
    \includegraphics[width=19pc,height=14pc]{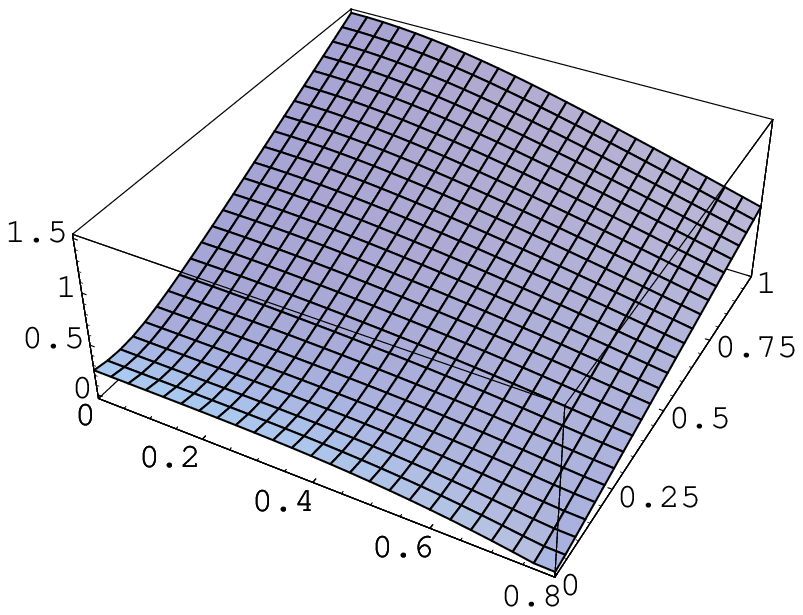}~~\quad
   \includegraphics[width=19pc,height=14pc]{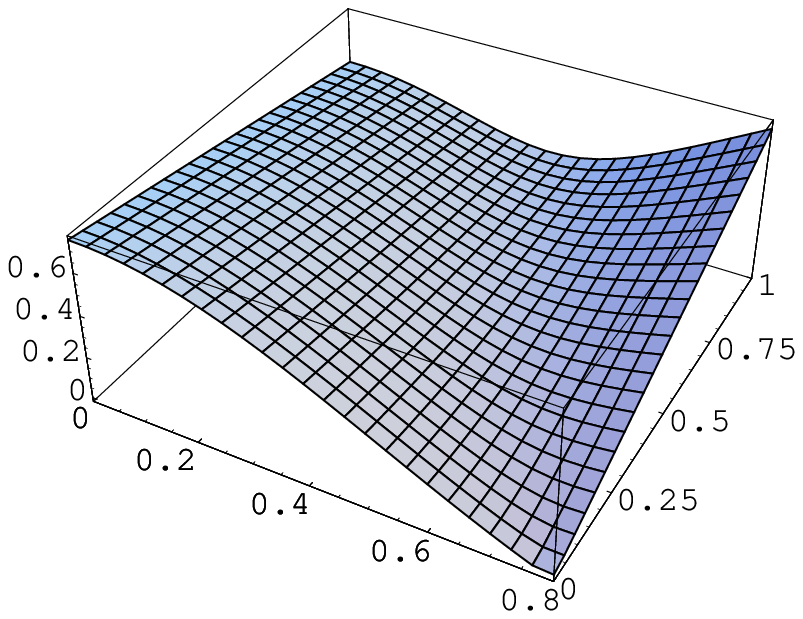}
 \put(-490,80){$Telp$}
 \put(-245,75){$Telp$}
\put(-150,20){$r$} \put(-260,50){$x$}
 \put(-400,20){$r$}
\put(-20,50){$p$}
 \put(-420,150){$(a)$}
 \put(-28,150){$(b)$}\\
   \includegraphics[width=19pc,height=14pc]{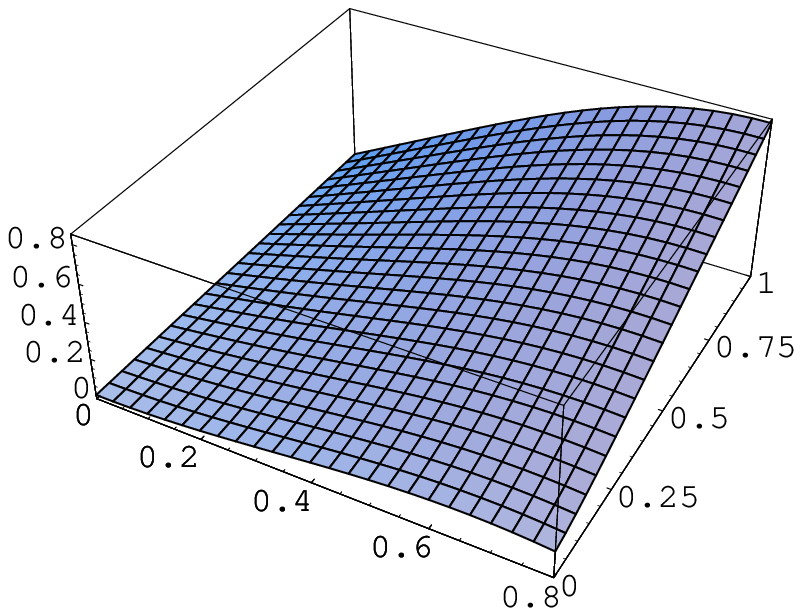}~\quad
   \includegraphics[width=19pc,height=14pc]{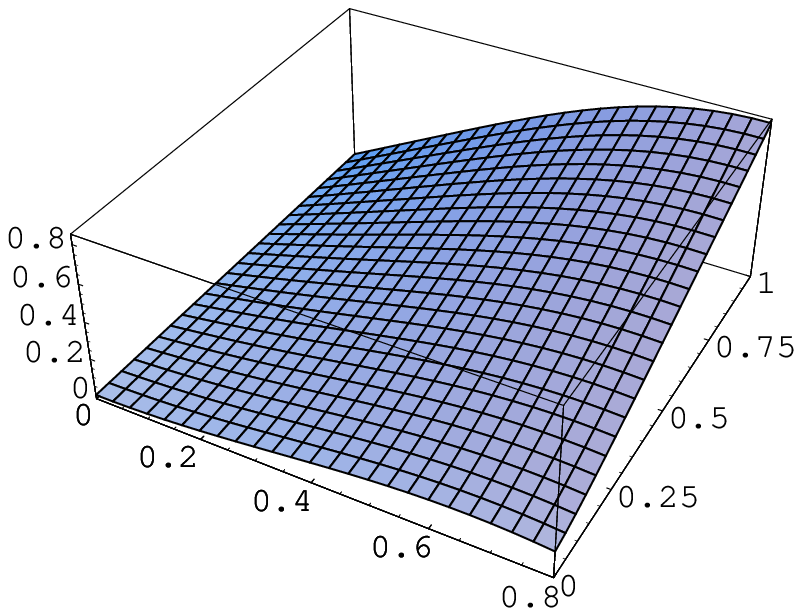}
\put(-420,150){$(c)$}
 \put(-28,150){$(d)$}
 \put(-490,80){$Telp$}
 \put(-245,75){$Telp$}
 \put(-150,20){$r$} \put(-260,50){$x$}
 \put(-400,20){$r$}
\put(-20,50){$p$}
     \caption{The teleportation inequality (\ref{Telp})  of the  travelling channel,
    where it is assumed that both particles are accelerated i.e., $r_a=r_b=r$ in the first
    region $I$ for a system  initially prepared in
   a self transposed type(MES or Werner) for (a)
   $\rho_{\tilde{A}_I\tilde{R}_I}$,
   (b)$\rho_{\tilde{A}_{II}\tilde{R}_{II}}$, (c) $\rho_{\tilde{A}_I\tilde{R}_{II}}$ and (d)
   $\rho_{\tilde{R}_I\tilde{A}_{II}}$.}
  \end{center}
\end{figure}
\\
 To examine wether the
travelling state can be used as a quantum channel to implement
teleportation, we use Hordecki's criterion \cite{Hor}. This
criterion state that any mixed spin $\frac{1}{2}$ state is useful
for quantum teleportation if
$tr\sqrt{~\dyadic{C}^T~\dyadic{C}~}>1.$ The self transposed states
are useful for quantum teleportation if,
\begin{equation}\label{Telp}
Telp=\sqrt{\tilde C^{2(i)}_{xx}+\tilde C_{yy}^{2(i)}+\tilde
C_{zz}^{2(i)}}>1,
\end{equation}
where $i=I,II$ for the first and second regions respectively.
Fig.(6), shows the possibility of using the travelling state as
quantum channel between two users to perform the quantum
teleportation protocol. In Fig.(\ref{Fig:tp}a), we consider that
the users initially share a maximum entangled state or Werner
state. it is clear that, for small values of $r$ and small values
of $x$ the teleportation inequality $\mathrm{Telp}<1$. However as
$x$ increases while the acceleration $r$ is small, the possibility
of using the channel for quantum teleportation  increases. For
larger values of $r$ and smaller values of  $x$, i.e., the system
is of Werner type with small degree of entanglement,
$\mathrm{Telp}<1$ and consequently the channel is useless for
quantum teleportation. As $x$ increases (we increases the degree
of entanglement), the quantum channel is useful for quantum
teleportation even for larger values of $r$.

 Fig.(6c) describes the behavior of the Horodecki's criterion
for the generated entangled channel between Alice and Anti-Rob. It
is clear that, the maximum value of the inequality (40) is smaller
than one ( $\mathrm{Telp}<1$) for all  self transposed classes.
Therefore according to   Horodecki's criterion, this channel is
useless for quantum teleportation. Also for the generated
entangled channel between Rob and Anti-Alice,  Horodeciki's
inequality is violated and consequently this class can not be used
as a quantum channel to perform quantum teleportation as shown in
Fig.(6d).

Fig.(7), displays the Horodecki's inequality (40) for a system
initially  prepared in the  generic pure state (29). It is clear
that for MES, namely $p=0$, the travelling channel can  be used as
a quantum channel to perform quantum teleportatiion $(Telp>1)$ for
$r\sim\in[0,4]$. However for larger values of $p\in[0,0.75]$ and
smaller values of $r$, the travelling channel obeys Horedicki's
inequality (\ref{Telp}) and consequently all these classes could
be used to implement the quantum teleportation protocol. Finally
the last classes which are characteristic by $p\in[0.75,1]$
violate the teleportation inequality (\ref{Telp}) and consequently
this classes could not be used as quantum channels to perform
teleportation.

\begin{figure}\label{tpp}
  \begin{center}
        \includegraphics[width=19pc,height=14pc]{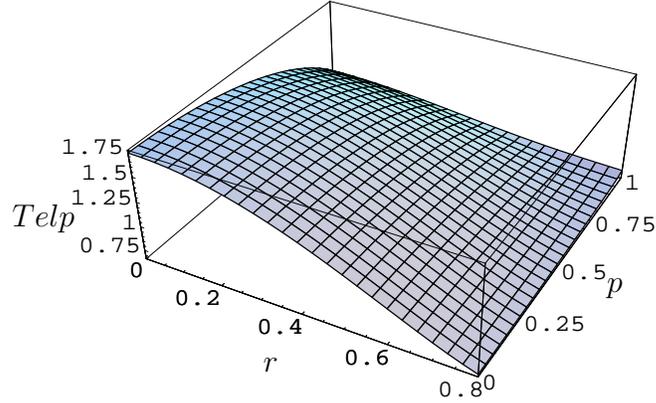}
   \put(-245,75){$Telp$}
\put(-150,20){$r$}
\put(-20,50){$p$}
    \caption{The teleportation inequality (\ref{Telp})  of the  travelling state ,
    where it is assumed that both particles are accelerated i.e., $r_a=r_b=r$ in the first
    region $I$  for a system prepared initially in a generic pure state (\ref{Pure}).}
  \end{center}
\end{figure}

From these results, one concludes that it is possible to generate
entangled channels  between Alice, Rob in the first and second
regions, Alice, Anti-Rob,  and between Rob, Anti-Alice. For MES,
the largest entanglement can  be generated between Alice and Rob
$\rho_{\tilde{A}_I\tilde{R}_I}$ in the first region and the
smallest values of entanglement contained in $\rho_{\tilde
A_{II}\tilde R_{II}}$. The generalized Werner state is more robust
than the MES and Werner classes. However the  pure states which
represent a class of partial entangled states are more robust than
the self transposed class.
 The possibility of using the pure state for quantum teleportation proposes are much better
than the other classes.
\end{enumerate}

\section{Conclusion}
 An analytical solution for  a general  two qubits system  of Dirac fields in noninertial frame is introduced.
 The density operator of the travelling channels between the observers and their Anti-observers are
calculated. Two particular classes are investigated extensively:
the  self transposed class, which includes the maximum entangled
state and all classes of Werner states and  a class of a  generic
pure state of a two qubits system .

For the self transposed states (MES or Werner) the degree of
entanglement in  all  different generated states depends on the
entanglement of the initial travelling state. The degree of
entanglement decreases smoothly if only one qubits accelerated and
it vanishes  completely  if the acceleration of at least one qubit
tends to infinity. However, the entanglement vanishes faster for
less initial entangled states. For a general class of generalized
Werner  state, the entanglement doesn't vanish even for larger
accelerations. Starting from a generic pure state, the
entanglement is more robust than that depicted for the self
transposed classes, where the entanglement decreases slowly and
gradually. For this class, we  show that,  an entangled channels
in the second region  with small degree of entanglement is
generated.

We distinguish between the usefulness of the travelling classes by
investigating  two phenomena:{\it first}, we quantify the fidelity
of the travelling  channels in the two regions and the {\it
second}, the possibility of using the travelling channels  as
quantum channels to perform the quantum teleportation. The
fidelity of the travelling channels depends on the entanglement of
the initial state, where it is larger for higher entanglement. For
the self transposed states, the fidelity is maximum for maximum
entangled state and decreases  smoothly as the acceleration of any
particle increases. For less entangled states, the fidelity is
smaller and decreases faster. In the second region, the fidelity
of the travelling  channel increases gradually and it is smaller
for the maximum entangled state comparing  with that in the first
region. On the other hand, the entanglement of the generated
entangled channels  between Alice and Anti-Rob is much larger than
that depicted between Alice and Rob in the second region.

The fidelity  of a travelling  channel starting from a generic
pure state is investigated in both regions, where the fidelity of
this class is much better than that depicted for the self
transposed classes. In the first region, the loss of the fidelity
is very small for the smallest entanglement of initial classes. On
the other hand, the fidelity of the generic pure state in the
second region is much better than its corresponding one for the
self transposed class.

The usefulness of the travelling  channels  is investigated for
different classes. We showed that it is possible to use some
classes of the self transposed states  for quantum teleportation.
These classes are generated from a system that has an initial
degree of entanglement ($>0.5$) and its qubits are accelerated
with an acceleration $<0.6$).  On the other hand, starting from a
generic pure state with smaller values of its parameter, one can
generate entangled channels in the first region useful for quantum
teleportation with larger acceleration

In {\it conclusion}, the possibility of generating entangled
channels in both regions depends on the degree of entanglement of
the initial travelling state. The fidelity and the usefulness of
the generic pure state is much better than that depicted for the
self transposed classes. Therefore, one can say that the generic
pure state is more robust than the self transposed classes.

\end{document}